\documentclass{PoS}

\usepackage{amsfonts} 
\usepackage{amssymb} 
\usepackage{amsmath} 
\usepackage{graphicx}    
\usepackage{subcaption}
\usepackage{multirow}
\usepackage[dvipsnames]{xcolor}

\providecommand{\fm}{\,\mathrm{fm}}

\newcommand{\CO}{{\cal O}}
\newcommand{\CR}{{\cal R}}
\newcommand{\kcrit}{\kappa_\text{crit}}

\newcommand{\BtoDst}{\bar B \rightarrow D^\ast \ell \bar \nu}

\newcommand{\epsK}{\varepsilon_{K}}

\newcommand{\cdf}{\chi^2/\text{dof}}

\title{Update on $B\to D^\ast \ell \nu$ form factor at zero-recoil
  using the Oktay-Kronfeld action}

\ShortTitle{$B\to D^\ast \ell \nu$ form factor at
  zero-recoil}

\author{Tanmoy Bhattacharya, Rajan Gupta, \speaker{Sungwoo Park}
  \\ Theoretical Division T-2, Los Alamos National Laboratory,
  Los Alamos, NM 87545, USA
  \\ E-mail: \email{tanmoy@lanl.gov, rg@lanl.gov, sungwoo@lanl.gov}}

\author{Yong-Chull Jang
  \\ Phsics Department, Brookhaven National Laboratory, Upton, NY 11973, USA
  \\ E-mail: \email{ypj@bnl.gov}}

\author{ Jon A. Bailey, Benjamin J. Choi, Hwancheol Jeong,
  Seungyeob Jwa, Sunkyu Lee, Weonjong Lee, Jeonghwan Pak\\
  Lattice Gauge Theory Research Center, CTP, and FPRD, \\
  Department of Physics and Astronomy,
  Seoul National University, Seoul 08826, South Korea\\
  E-mail: \email{jabsnu@gmail.com, wlee@snu.ac.kr}}

\author{ Jaehoon Leem \\
  School of Physics, Korea Institute for Advanced Study (KIAS),
  Seoul 02455, South Korea
  \\ E-mail: \email{leemjaehoon@kias.re.kr}}

\author{LANL/SWME Collaboration}

\abstract{We present an update on the calculation of $\bar{B}\to
  D^\ast \ell \bar{\nu}$ semileptonic form factor at zero recoil using
  the Oktay-Kronfeld bottom and charm quarks on $N_f=2+1+1$ flavor
  HISQ ensembles generated by the MILC collaboration. Preliminary
  results are given for two ensembles with $a\approx 0.12$ and $0.09$
  fm and $M_\pi\approx 310$ MeV. Calculations have been done with a
  number of valence quark masses, and the dependence of the form
  factor on them is investigated on the $a\approx 0.12\,\fm$
  ensemble. The excited state is controlled by using multistate fits
  to the three-point correlators measured at 4--6 source-sink
  separations. }

\FullConference{The 36th Annual International Symposium on Lattice
  Field Theory - LATTICE2018\\ 22-28 July, 2018\\ Michigan State
  University, East Lansing, Michigan, USA.}

\begin{document}

\section{Introduction}

Lattice calculation of $\BtoDst$ semileptonic decay form factors can
be used to determine the CKM matrix element $|V_{cb}|$ from the
measured exclusive decay rates. Precise results for exclusive
$|V_{cb}|$ will address the well-known $\approx 3\sigma$ discrepancy
from the inclusive determination of
$|V_{cb}|$~\cite{Amhis:2016xyh}. In addition, there is an analysis
showing $\approx 4\sigma$ tension in the standard model evaluation of
$|\epsK|$~\cite{Bailey:2018feb} using the exclusive $|V_{cb}|$.

For the $\BtoDst$ study, the heavy quark discretization error, 
estimated by HQET power counting in terms of 
$\displaystyle \lambda_c\sim \frac{\Lambda_{QCD}}{2m_c}\sim
\frac{500~\text{MeV}}{2\times 1.3~\text{GeV}}\sim \frac{1}{5}$, is
dominant especially for charm.
Calculations of the zero recoil form factor
$\mathcal{F}(w=1)=h_{A_1}(1)$ using the Fermilab action has
$\CO(\lambda^3_c)\sim 1\%$ uncertainty assuming $\alpha_s \sim \lambda_c$. To achieve
precision below 1\% in $|V_{cb}|$, we propose to use the
Oktay-Kronfeld (OK) action, in which the discretization error are 
$\CO(\lambda^4_c) \sim 0.2\%$ provided a full one-loop improvement of correction terms is 
carried out~\cite{Oktay:2008ex, Bailey:2017nzm}. In this work, we are working 
with tree-level tadpole-improvement of the action and current operators since the 
one-loop calculations are not complete.

We calculate the form factor $h_{A_1}(1)$ using the double ratio of ground state matrix elements \cite{Bailey:2014tva},
\begin{align}
  |{h_{A_1}(1)}|^2 = \frac{ \langle D^\ast
    |A^j_{cb}| \bar{B}\rangle \langle \bar{B} |A^j_{bc}|
    D^\ast\rangle}{\langle D^\ast |V^4_{cc}| D^\ast\rangle\langle
    \bar{B} |V^4_{bb}| \bar{B}\rangle} \times \rho_{A_j}^2\,,
  \quad {\rm with} \quad 
    \rho^2_{A_j} = \frac{Z_{A_j}^{cb}Z_{A_j}^{bc}}{Z_{V_4}^{cc}Z_{V_4}^{bb}} \,.
  \label{eq:h_A1}
\end{align}
$\rho^2_{A_j}$ is the matching factor that is expected to be
close to unity~\cite{Bailey:2014tva}.  Each of the matrix element is
extracted from the related three-point function calculated on the
lattice. For example, $\langle D^\ast |A^j_{cb}| B\rangle$ is from
$C^{B\rightarrow D^\ast}_{A_1}(t,\tau)$ defined in
Eq.~\eqref{eq:C_B2Dstar}.
%
%
We also use an improved current operator $A_j^{cb}(y)=\bar{\Psi}_c(y)
\gamma_j \gamma_5 \Psi_b(y)$ where the improved field $\Psi (x)$ is
obtained by the following field rotation on the unimproved fermion
field $\psi(x)$:
%
$  \Psi (x) = [1+\sum_{i}d_i\CR_i]\psi(x)$. 
Improvement up to $\CO (\lambda^3)$ can be obtained using tree-level
matching of the coefficients $d_i$ and operators
$\CR_i$~\cite{ElKhadra:1996mp, Bailey:2017zgt}.

We present an update on our previous calculation of
$|h_{A_1}(1)/\rho_{A_j}|$ \cite{Bailey:2017xjk}, paying attention to
controlling the excited-state contamination (ESC) using multistate
fits to 3-point (3pt) data with 4--6 source-sink time separations.
The parameters in the calculation on the two $N_f=2+1+1$ HISQ ensembles, generated by the
MILC collaboration~\cite{Bazavov:2012xda}, are given in Table
\ref{tab:param}. The light quark propagators ($u,d,s$) are calculated
using the HISQ action with point source and sink. For the heavy quark
($c,b$) propagators, we use the OK action with covariant Gaussian
smearing at both the source and the sink. The hopping parameters
$\kcrit$, $\kappa_b$, $\kappa_c$ have been tuned nonperturbatively
as described in Ref.~\cite{Bailey:2017xjk}.\looseness-1
\begin{table}[t]
  \vspace{-5mm}
  \center
  \resizebox{1.0\textwidth}{!}{%
  \begin{tabular}{c|cccccccccc}
    \hline\hline
    ID  & $a$  & $M_\pi$ & $m_x/m_s$ & $\kappa_\text{crit}$ & $\kappa_c$ & $\kappa_b$ & $\{\sigma,N_\text{cvg}\}$ & $N_\text{cfg}\times N_\text{src}$ & $\tau$ \\\hline
    \multirow{2}{*}{$a12m310$} & \multirow{2}{*}{$ 0.1207$} & \multirow{2}{*}{305} & 0.1, $0.2^\dagger$, & \multirow{2}{*}{0.051211} & \multirow{2}{*}{0.048524} & \multirow{2}{*}{0.04102} & \multirow{2}{*}{$\{1.5,5\}$} &  \multirow{2}{*}{$1053\times 3$} & 10, 11, 12, \\
    &  &  & 0.3, 0.4, 1.0 & & & & & & 13, 14, 15 \\\hline
    $a09m310$ & {$ 0.0888$} & {313} & $0.2^\dagger$, 1.0 & 0.05075 & 0.04894 & 0.0429  & $\{2,10\}$ & $1001\times 3$ & 15, 16, 17, 18 \\
    \hline\hline
  \end{tabular}}
  
    \caption{Parameters used in the measurements performed on two MILC
      HISQ gauge ensembles described in
      Ref.~\cite{Bazavov:2012xda}. $m_x/m_s$ is the ratio of valence
      spectator quark mass to the sea strange quark mass where
      $\dagger$ denotes the unitary point for the degenerate up and
      down quarks. Hopping parameters $\kappa_\text{crit}$, $\kappa_c$
      and $\kappa_b$ give the values obtained for the critical and the
      charm and the bottom quark masses.  $\{\sigma,N_\text{cvg}\}$
      are parameters for the covariant Gaussian
      smearing. $N_\text{cfg}\times N_\text{src}$ denotes the number
      of measurements made. $\tau$ gives the source-sink time
      separations simulated.  }
  \label{tab:param}
\end{table}

\section{Controlling excited-state contamination}
\begin{figure}[t]
  \vspace{-0mm}
\centering
\begin{subfigure}[t]{.47\textwidth}
  \vspace{0pt}
  \includegraphics[width=1\linewidth]{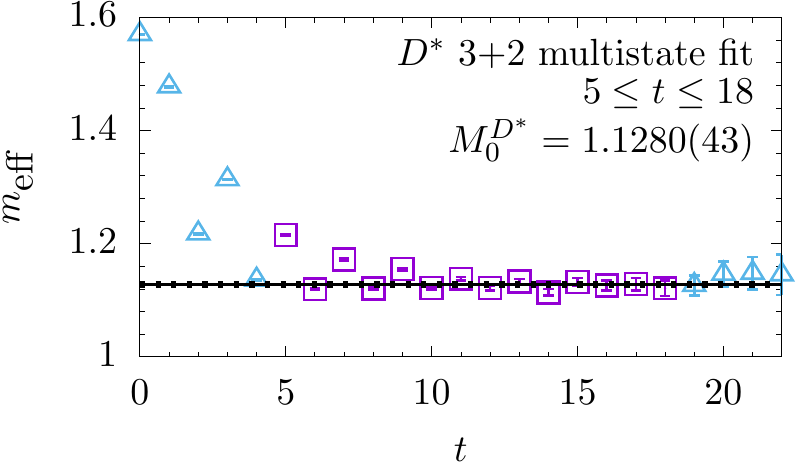}
\end{subfigure}
\hfill
\begin{subfigure}[t]{.42\textwidth}
  \vspace{0pt}
  \includegraphics[width=1\linewidth]{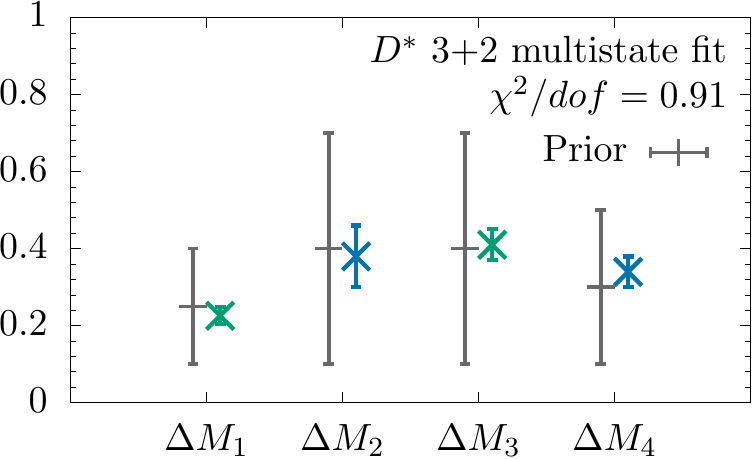}
\end{subfigure}
  \caption{(Left) Effective mass, $ m_\text{eff}(t)\equiv \frac{1}{2}\ln |    C^{2pt}(t)/C^{2pt}(t+2)|$,  plot 
   using the 3+2-state fit to the 2pt function on the a12m310 ensemble for 
   the $D^\ast$-meson with $m_x/m_s=0.2$. The ground state mass is shown as a horizontal
    line. (Right) The excited state masses from the 3+2-states 
    fit using empirical Bayesian priors. }
  \label{fig:2pt}
\end{figure}

To achieve sub-percent precision, we have to control the ESC. 
On a lattice with time extent $T$, the $B$- and $D^\ast$-meson
2-point (2pt) functions, $C^\text{2pt}(t)$, are fit using a $3+2$-state ansatz:
\begin{align}
  C^\text{2pt}(t)=\langle O^\dagger(t) O(0) \rangle &={|\mathcal{A}_0|^2} e^{-{M_0}
    t}\Big( 1
  +{\left|\frac{\mathcal{A}_2}{\mathcal{A}_0}\right|^2}e^{-{ \Delta M_2} t}
  +{\left|\frac{\mathcal{A}_4}{\mathcal{A}_0}\right|^2}
  e^{-{ (\Delta M_2 + \Delta M_4)} t}+ \cdots \\
  & -(-1)^t{\left|\frac{\mathcal{A}_1}{\mathcal{A}_0}\right|^2}
  e^{-{\Delta M_1}t}
  -(-1)^t {\left|\frac{\mathcal{A}_3}{\mathcal{A}_0} \right|^2}
  e^{-{(\Delta M_1 + \Delta M_3)} t}+\cdots \Big) +(t\leftrightarrow T-t).\nonumber
\end{align}
where $O$ is the meson interpolating operator. $\Delta M_{n} \equiv
M_n -M_{n-2}$ with $n=2,4$ are the mass gaps for even parity,
and $\Delta M_{1} \equiv M_1 -M_{0}$ and $\Delta M_{3} \equiv M_3
-M_{1}$ for the two odd parity states that arise in staggered
formulations.  An empirical Bayesian method is used to fix the priors
for the excited-state masses $M_n$ and amplitudes
$\mathcal{A}_n=\langle n | O| \Omega \rangle $ to stabilize the fits
as described in Ref.~\cite{Yoon:2016jzj}.
Fig.~\ref{fig:2pt} illustrates the results for the ground- and
excited-state masses of $D^\ast$ meson and the priors used for excited
states.

\begin{figure}[p]
\center
  \includegraphics[width=0.49\textwidth]{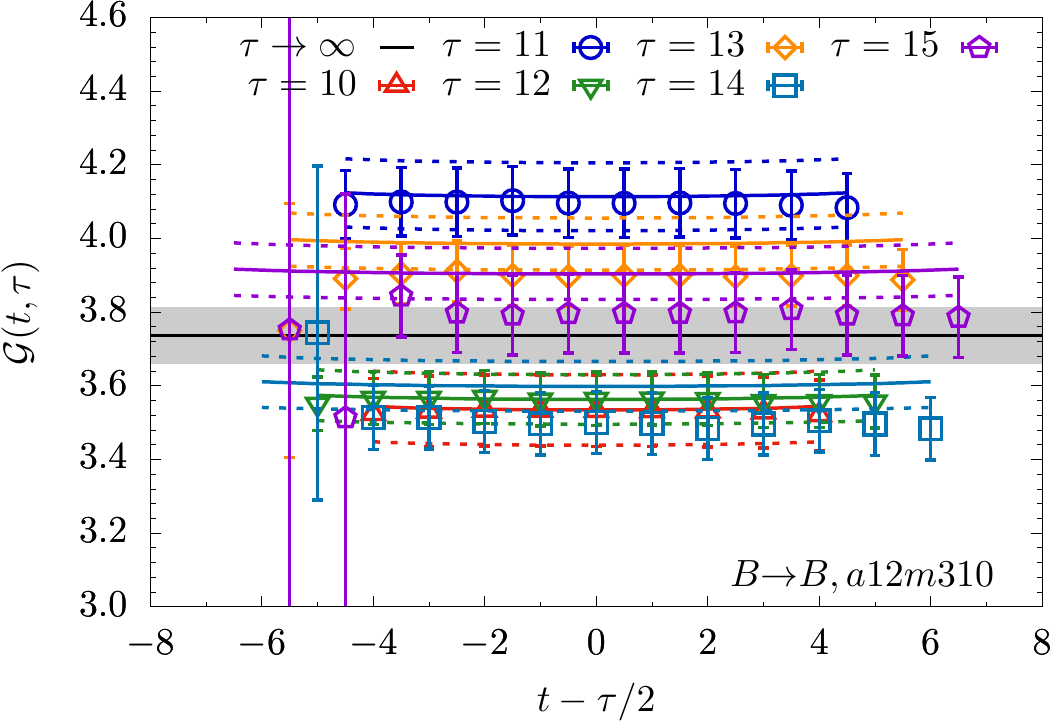}
  \includegraphics[width=0.49\textwidth]{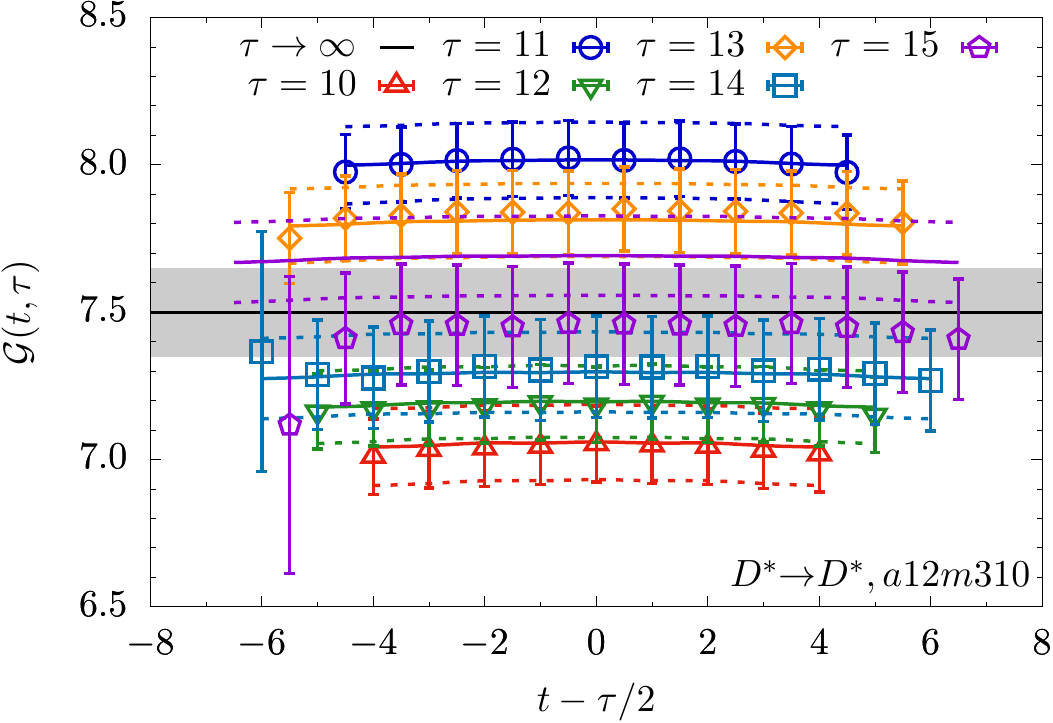}
  \includegraphics[width=0.49\textwidth]{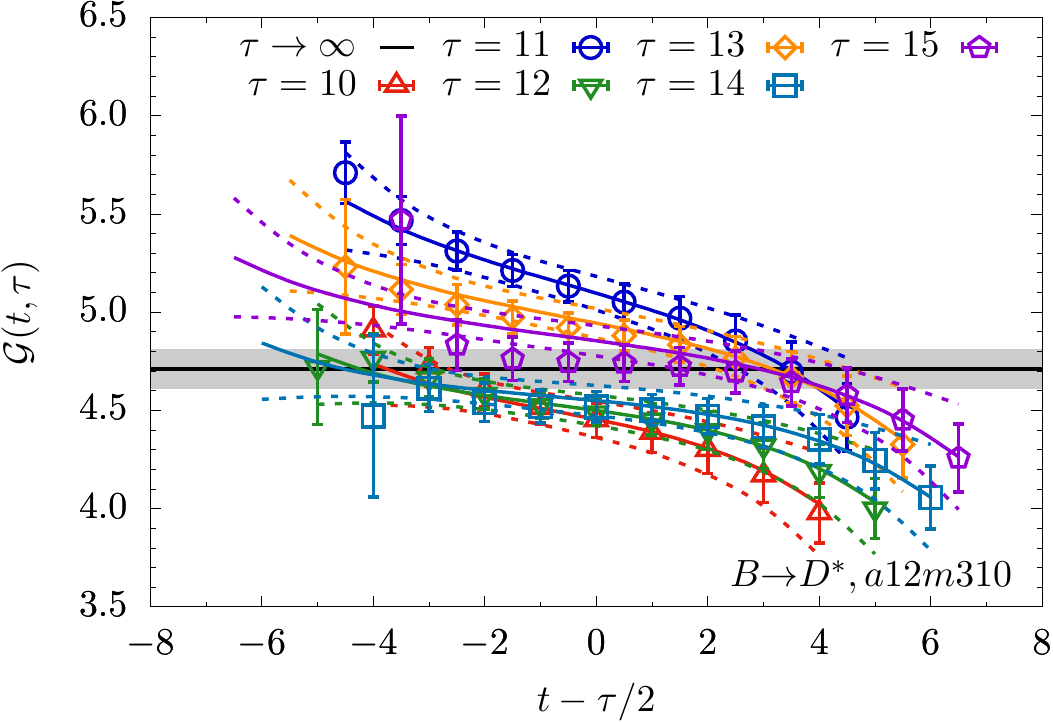}
  \includegraphics[width=0.49\textwidth]{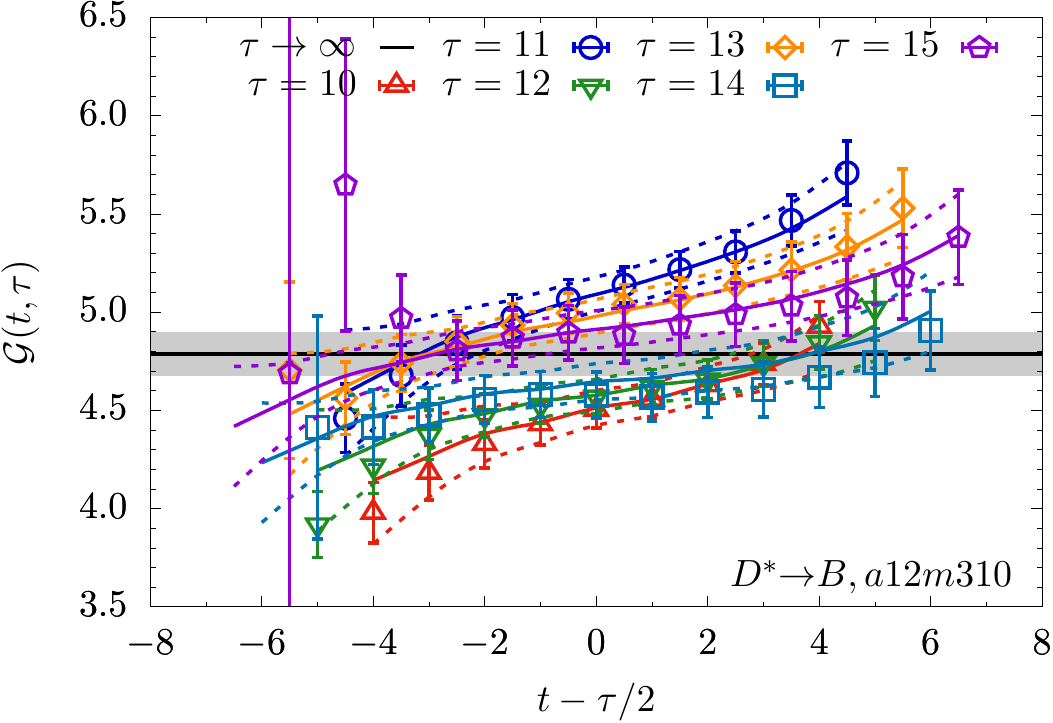}
  \includegraphics[width=0.49\textwidth]{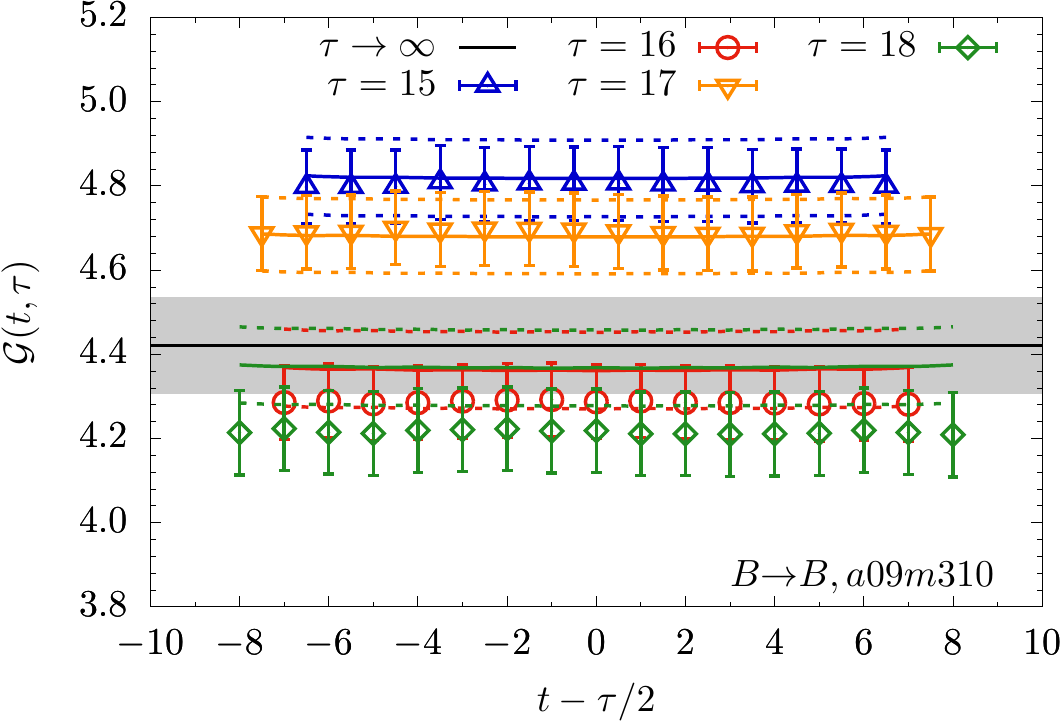}
  \includegraphics[width=0.49\textwidth]{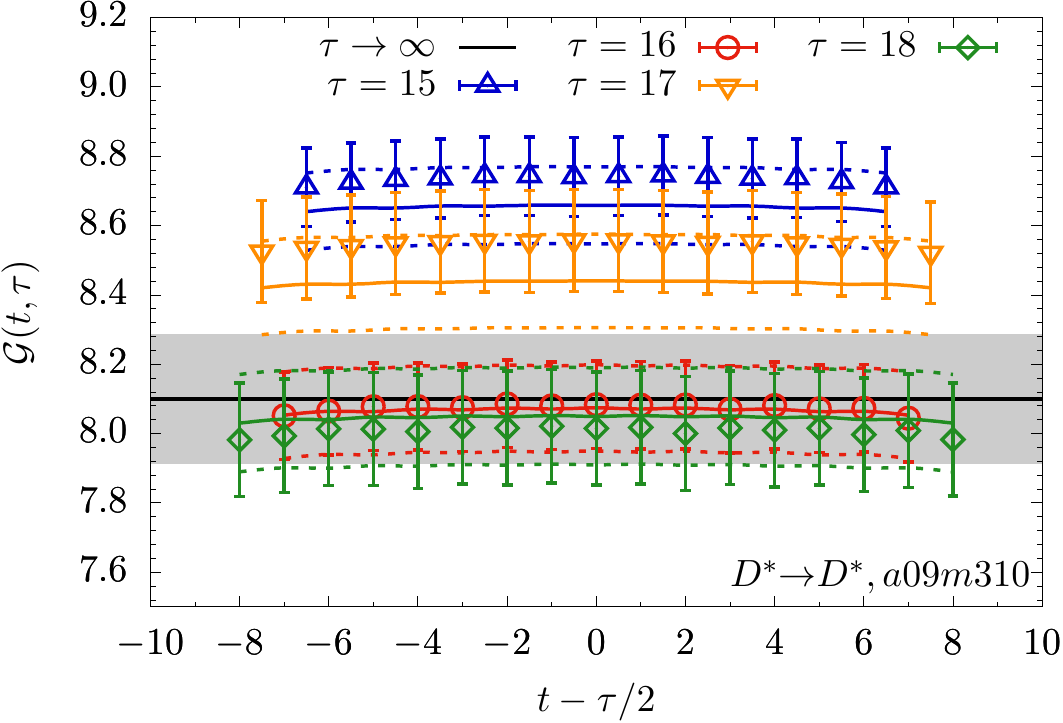}
  \includegraphics[width=0.49\textwidth]{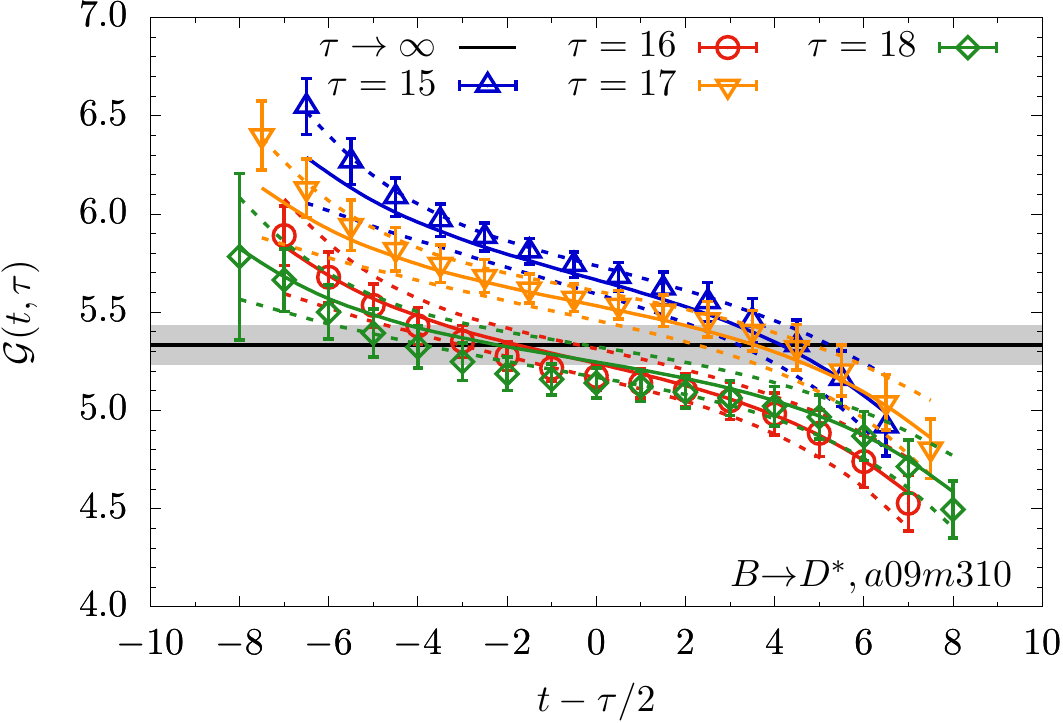}
  \includegraphics[width=0.49\textwidth]{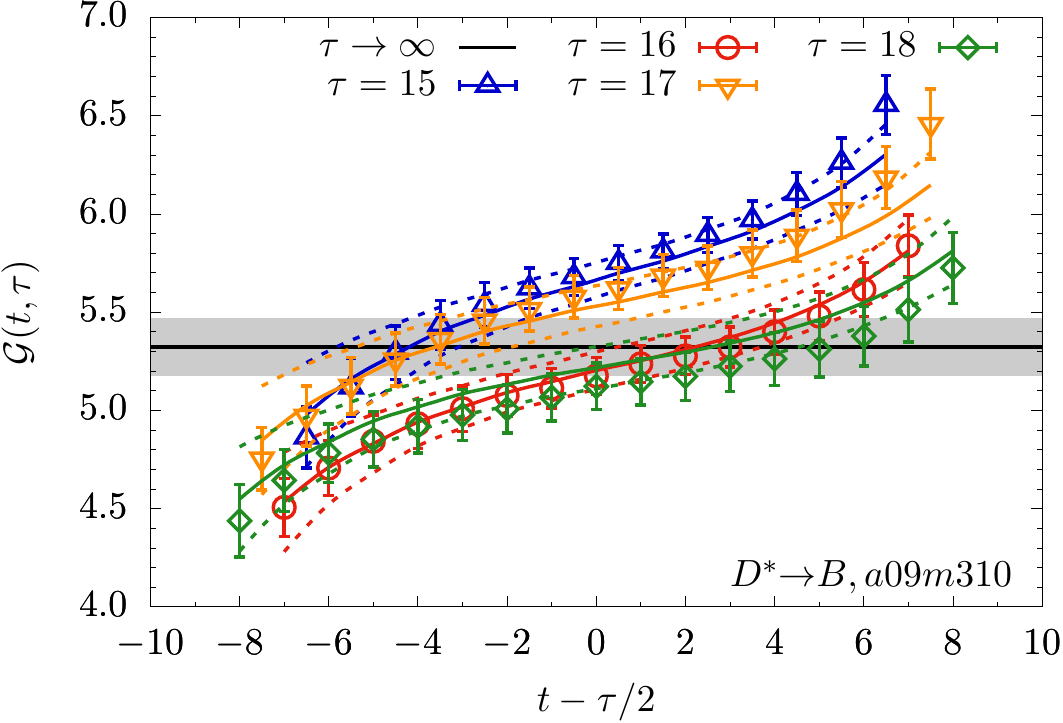}
  \caption{Data for the ratio $\mathcal{G}$ for the various values of
    $\tau$ (see labels) are plotted versus $t-\tau/2$ for the
    $\CO(\lambda^2)$ improved current and $m_x=0.2m_s$. The horizontal
    line is the ground-state matrix element ($\tau \to \infty$) determined from the
    multistate fit using Eq.~\protect\eqref{eq:spectral}. The results
    of the fit for each $\tau$ is shown in the same color as the data.
    Note the difference in ESC for $D^\ast\rightarrow B$ (or the rough mirror process 
    $B\rightarrow D^\ast$) and $D^\ast\rightarrow D^\ast$ (or $B\rightarrow B$) and 
    the change with $a$ between $a12m310$ (top 4 panels) and $a09m310$ (bottom 4 panels). 
}
  \label{fig:ratio}
\end{figure}

The 3pt data is fit including 
$2+1$ states for $|B_m \rangle $ and $|D^\ast_n \rangle$
in the spectral decomposition: \looseness-1
\begin{align}
    &C^{B\rightarrow D^\ast}_{A_j}(t,\tau)= \langle O_{D^\ast}^\dagger(0)
  A_j^{cb}(t)O_{B}(\tau) \rangle \qquad (0<t<\tau)
  \label{eq:C_B2Dstar}\\
  &= {\mathcal{A}^{D^\ast}_0 \mathcal{A}^{B}_0}
  {\langle D^\ast_0 | A^{cb}_j | B_0\rangle}
  e^{-{M_{B_0}} (\tau-t)}  e^{-{M_{D^\ast_0}} t}  - {\mathcal{A}^{D^\ast}_0 \mathcal{A}^{B}_1}
  {\langle D^\ast_0 |  A^{cb}_j | B_1\rangle}
  (-1)^{\tau-t} e^{-{M_{B_1}} (\tau-t)}  e^{-{M_{D^\ast_0}} t} \nonumber\\
  &\quad  - {\mathcal{A}^{D^\ast}_1 \mathcal{A}^{B}_0}
  {\langle D^\ast_1 | A^{cb}_j | B_0\rangle}
  (-1)^{t}e^{-{M_{B_0}} (\tau-t)}  e^{-{M_{D^\ast_1}} t}  + {\mathcal{A}^{D^\ast}_1 \mathcal{A}^{B}_1}
  {\langle D^\ast_1 |   A^{cb}_j | B_1\rangle}
  (-1)^{\tau}e^{-{M_{B_1}} (\tau-t)}  e^{-{M_{D^\ast_1}} t} \nonumber\\
  &\quad  + {\mathcal{A}^{D^\ast}_2 \mathcal{A}^{B}_0}
  {\langle D^\ast_2 | A^{cb}_j | B_0\rangle}
 e^{-{M_{B_0}} (\tau-t)}  e^{-{M_{D^\ast_2}} t}   + {\mathcal{A}^{D^\ast}_0 \mathcal{A}^{B}_2}
  {\langle D^\ast_0 | A^{cb}_j | B_2\rangle}
  e^{-{M_{B_2}} (\tau-t)}  e^{-{M_{D^\ast_0}} t}\nonumber\\
  &\quad  - {\mathcal{A}^{D^\ast}_2 \mathcal{A}^{B}_1}
  {\langle D^\ast_2 | A^{cb}_j | B_1\rangle}
  (-1)^{\tau-t} e^{-{M_{B_1}} (\tau-t)}  e^{-{M_{D^\ast_2}} t}
   - {\mathcal{A}^{D^\ast}_1 \mathcal{A}^{B}_2}
  {\langle D^\ast_1 | A^{cb}_j | B_2\rangle}
  (-1)^{t} e^{-{M_{B_2}} (\tau-t)}  e^{-{M_{D^\ast_1}} t}\nonumber\\
  &\quad  + {\mathcal{A}^{D^\ast}_2 \mathcal{A}^{B}_2}
  {\langle D^\ast_2 | A^{cb}_j | B_2\rangle}
  e^{-{M_{B_2}} (\tau-t)}  e^{-{M_{D^\ast_2}} t}  +\cdots,
  \label{eq:spectral}
\end{align}
%
%
where $A_j^{cb}$ is the improved axial current inserted at time $t$,
and $\mathcal{A}^{B}_n$, $\mathcal{A}^{D^\ast}_m$, $M_{D^\ast_n}$ and
$M_{B_m}$ values are taken from fits to the 2pt functions. 
Similar fit functions are used for the other channels: $C^{D^\ast\rightarrow
  B}_{A_1}(t,\tau)$, $C^{B\rightarrow B}_{V_4}(t,\tau)$ and
$C^{D^\ast\rightarrow D^\ast}_{V_4}(t,\tau)$. 
In all the fits, we skip four points next to the source and the sink that have the largest ESC. \looseness-1
In Fig.~\ref{fig:ratio}, we display the ratio, $\mathcal{G}$,
\begin{align}
  \mathcal{G}(t,\tau) &\equiv \frac{C^{B\to
      {D^\ast}}_{A_j}(t,\tau)}{\mathcal{A}^{D^\ast}_0
    \mathcal{A}^{B}_0e^{-M_{B_0}(\tau-t)} e^{-M_{D^\ast_0} t} }=
  \langle {D^\ast_0} | A_j^{cb} | B_0\rangle + \cdots, 
  \label{eq:3pt-to-2pt}
\end{align}
where $\mathcal{A}^{B}_0$, $\mathcal{A}^{D^\ast}_0$, $M_{D^\ast_0}$
and $M_{B_0}$ are ground-state amplitudes and masses determined from
2pt function fits. $\mathcal{G}$ asymptotes to the ground state matrix
element in the $t\to\infty$ and $\tau-t\to \infty$ limits. The data
show the size of the ESC and the oscillatory nature of the
convergence. The $(-1)^\tau{\langle D^\ast_1 | A^{cb}_j | B_1\rangle}$
term controls the even-odd oscillation about the grey band while
${\langle D^\ast_2 | A^{cb}_j | B_2\rangle}$ controls the convergence
as $\tau \to \infty$ for both even or odd $\tau$ data. We also find
that the contribution of terms of the form $(-1)^{\tau-t}{\langle D^\ast_0 | A^{cb}_j |
  B_1\rangle}$ is tiny and that of $(-1)^{\tau-t}{\langle D^\ast_1 |
  A^{cb}_j | B_2\rangle}$ is negligible. The latter is therefore set to zero in the final
fits.  On the other hand, the grey horizontal band in
Fig.~\ref{fig:ratio} is the ground-state matrix element obtained by
fitting $C^{B\to {D^\ast}}_{A_j}(t,\tau)$ using
Eq.~\eqref{eq:spectral}, to which $\mathcal{G}$ should converge.
These results from the fits are summarized in
Table~\ref{tab:mat_elem}.
Note that there is no significant improvement at $\CO(\lambda)$ in
$\langle B|V_4| B \rangle$ and $\langle D^\ast |V_4| D^\ast \rangle$,
but a large change at $\CO(\lambda^2)$~\footnote{Here, we used the 
    $\mathcal{O}(\lambda^2)$ improvement coefficient for the current
    given in the Ref.~\cite{ElKhadra:1996mp}. Taking the coefficient
    from Ref.~\cite{Bailey:2017zgt} results in a negligible change. Full
    $\mathcal{O}(\lambda^3)$ current improvement presented in
    Ref.~\cite{Bailey:2017zgt} is being implemented.}  (also see
Fig.~\ref{fig:hA1}).  Thus, higher order improvements for these two
channels may be necessary.\looseness-1

\begin{table}[t]
  \vspace{-5mm}
  \begin{subtable}[b]{\linewidth}
    \resizebox{\textwidth}{!}{
    \begin{tabular}{c |cl|cl|cl|cl}
      \hline\hline
      Current &    $\langle B | V_4 | B \rangle $ &       $\cdf\;[p]$   &      $\langle D^\ast | V_4 | D^\ast \rangle $  &       $\cdf\;[p]$ &      $\langle D^\ast | A_j | B \rangle $  &       $\cdf\;[p]$ &  $\langle B | A_j | D^\ast \rangle $  &       $\cdf\;[p]$ \\ \hline
                 Unimp. & 3.88(8)      & 1.20\;[0.21]  & 8.44(17)     & 0.73\;[0.84]  & 4.79(10)     & 0.73\;[0.84]  & 4.84(11)     & 0.89\;[0.63]  \\ 
      $\CO(\lambda)$ & 3.89(8)      & 1.20\;[0.21]  & 8.45(17)     & 0.73\;[0.84]  & 4.90(10)     & 0.74\;[0.83]  & 4.96(12)     & 0.84\;[0.70]  \\ 
      $\CO(\lambda^2)$ & 3.74(8)      & 1.27\;[0.15]  & 7.50(15)     & 0.86\;[0.68]  & 4.71(10)     & 0.73\;[0.84]  & 4.79(11)     & 0.93\;[0.56]  \\ 
      \hline\hline
    \end{tabular}}
  \caption{Results for the $a12m310$ ensemble with  $m_x=0.2m_s$.}
  \end{subtable}

  \vspace{1mm}

  \begin{subtable}[b]{\linewidth}
    \resizebox{\textwidth}{!}{
    \begin{tabular}{c |cl|cl|cl|cl}
      \hline\hline
      Current &    $\langle B | V_4 | B \rangle $ &       $\cdf\;[p]$   &      $\langle D^\ast | V_4 | D^\ast \rangle $  &       $\cdf\;[p]$ &      $\langle D^\ast | A_j | B \rangle $  &       $\cdf\;[p]$ &  $\langle B | A_j | D^\ast \rangle $  &       $\cdf\;[p]$ \\ \hline
                 Unimp. & 4.64(12)     & 1.44\;[0.05]  & 9.26(22)     & 0.90\;[0.63]  & 5.49(10)     & 1.54\;[0.03]  & 5.48(15)     & 0.68\;[0.91]  \\ 
      $\CO(\lambda)$ & 4.65(12)     & 1.44\;[0.05]  & 9.27(22)     & 0.90\;[0.63]  & 5.60(10)     & 1.54\;[0.03]  & 5.60(15)     & 0.67\;[0.92]  \\ 
      $\CO(\lambda^2)$ & 4.42(12)     & 1.34\;[0.09]  & 8.10(19)     & 0.84\;[0.73]  & 5.33(10)     & 1.62\;[0.02]  & 5.32(15)     & 0.76\;[0.83]  \\ 
      \hline\hline
    \end{tabular}}
  \caption{Results for the $a09m310$ ensemble with $m_x=0.2m_s$.}
  \end{subtable}
  \caption{Matrix elements of $\CO(\lambda^\ell)$ (with $\ell \in 0, 1, 2$) improved currents extracted from $2+1$-state fits.}
  \label{tab:mat_elem}
\end{table}

\section{$|h_{A_1}(1)/\rho_{A_j}|$ result}

\begin{figure}[t]
  \vspace{-5mm}
  \center
  \includegraphics[width=0.49\textwidth]{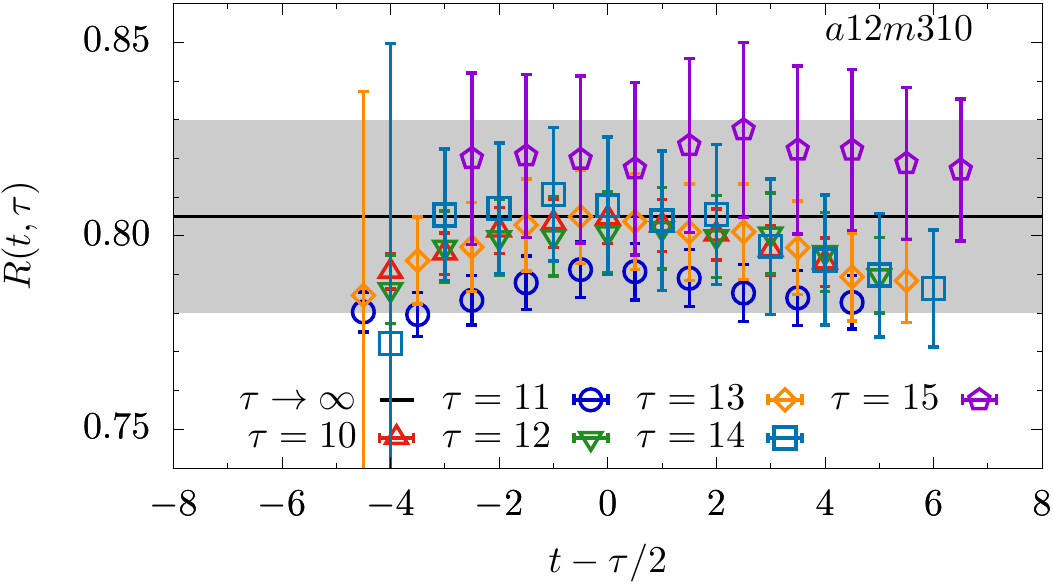}
  \includegraphics[width=0.49\textwidth]{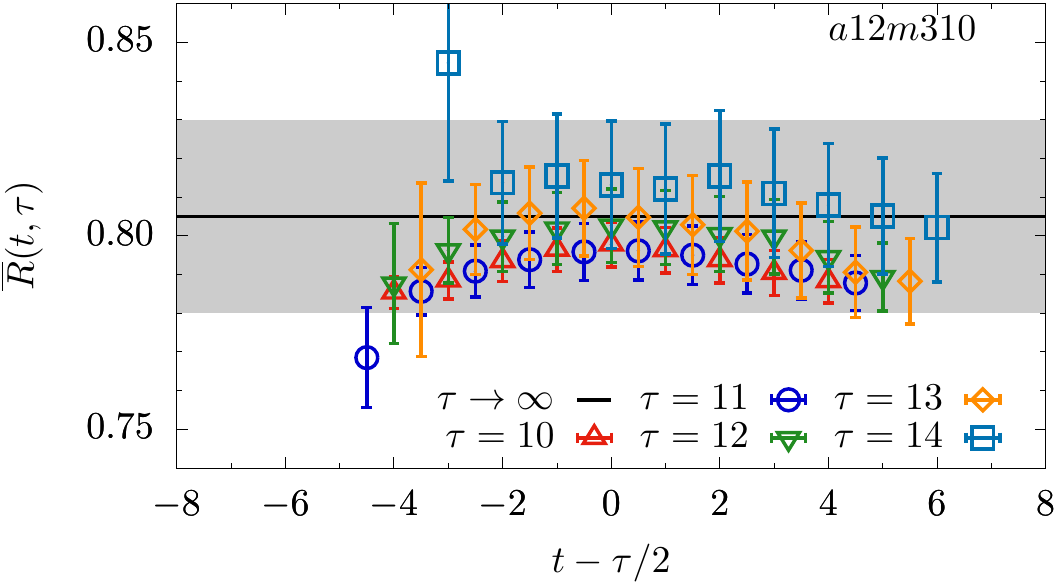}
  \includegraphics[width=0.49\textwidth]{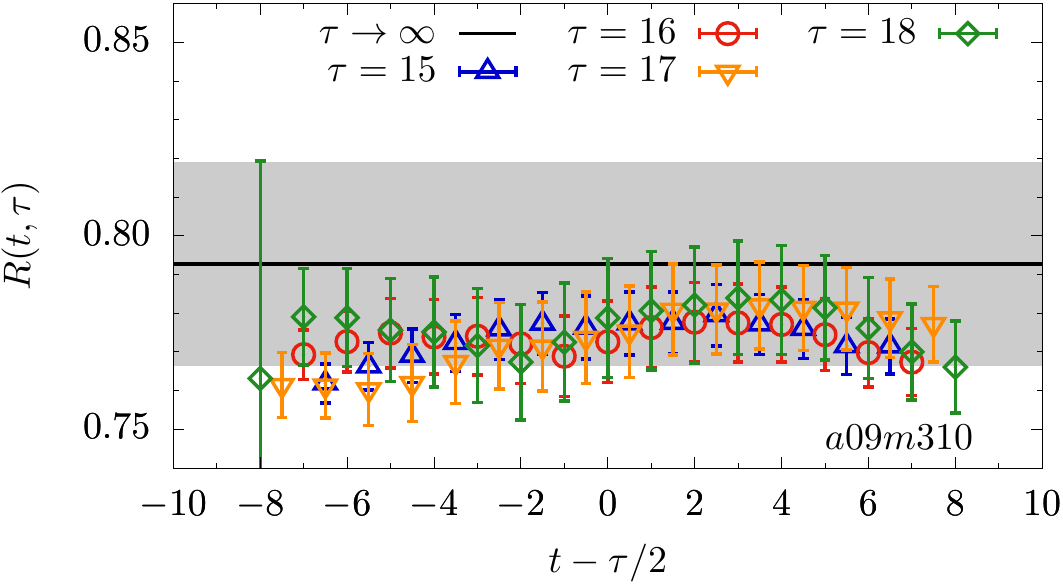}
  \includegraphics[width=0.49\textwidth]{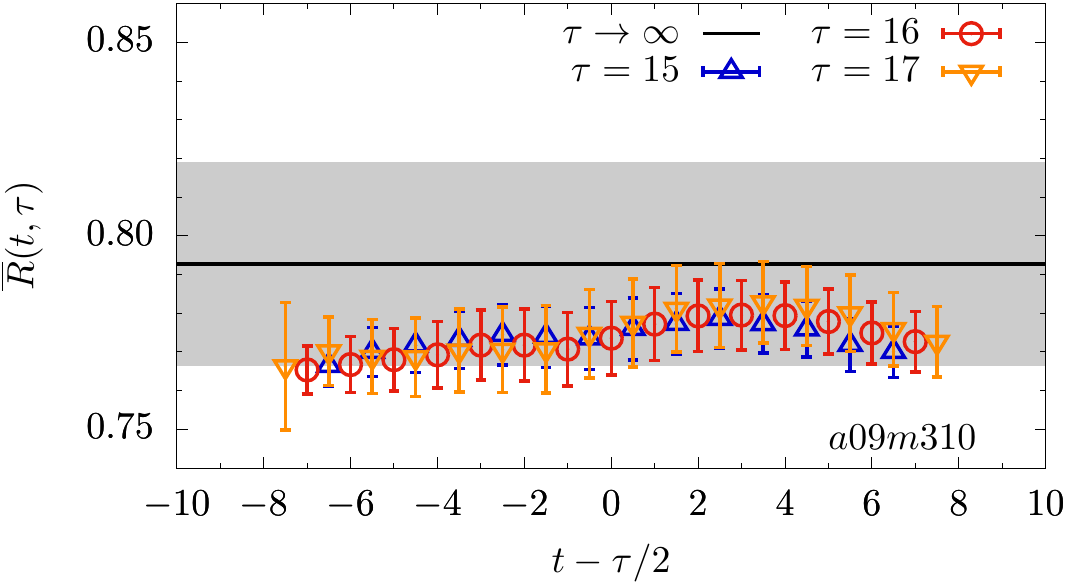}
  \caption{Data for the double ratio $R(t,\tau)$ for the various
    values of $\tau$ are plotted versus the operator insertion time
    $t- \tau/2$. The grey horizontal band is the result for
    $|h_{A_1}(1)/\rho_{A_j}|^2$ obtained using
    Eq.~\protect\eqref{eq:h_A1}.  The top (bottom) panels show data
    for the $a12m310$ ($a09m310$) ensemble at the unitary point $m_x=0.2m_s$. The
    left (right) panels show data for $R(t,\tau) $ ($
    \overline{R}(t,\tau)$) that are defined in the text.}
  \label{figs:R_a12}
\end{figure}

We obtain $|h_{A_1}(1)/\rho_{A_j}|^2$, defined in Eq.~\eqref{eq:h_A1}, using the ground-state matrix
elements given in Table~\ref{tab:mat_elem}. 
The result for the two values of the lattice spacing at fixed pion mass
$M_\pi \approx 310$~MeV (unitary point) is shown by the grey horizontal band in
all panels of Fig.~\ref{figs:R_a12}. The error estimate includes the uncertainty coming from the
fits used to remove excited-state effects.  The left two panels in Fig.~\ref{figs:R_a12} show 
the data for the double ratio $R(t,\tau)$
\begin{align}
  R(t,\tau) = \frac{C^{B\rightarrow
      D^\ast}_{A_1}(t,\tau)C^{D^\ast\rightarrow
      B}_{A_1}(t,\tau)}{C^{B\rightarrow B}_{V_4}(t,\tau)
    C^{D^\ast\rightarrow D^\ast}_{V_4}(t,\tau)},
    \label{eq:dbl-ratio}
\end{align}
that significantly cancels the ESC in each individual correlator
illustrated in Fig.~\ref{fig:ratio}. The right panels in
Fig.~\ref{figs:R_a12} show the linear combination
$\overline{R}(t,\tau)$ defined as~\cite{Bailey:2014tva},
\begin{align}
  \overline{R}(t,\tau) = \frac{1}{2}R(t,\tau)+\frac{1}{4}R(t,\tau+1)+\frac{1}{4}R(t+1,\tau+1) \,, 
    \label{eq:dbl-ratio-bar}
\end{align}
that further suppresses the ESC, especially from the opposite parity
states.  Since the grey band in Fig.~\ref{figs:R_a12} is constructed
as the ratio of ground state matrix elements (albeit evaluated using
2+1 state fits), both the ratios, $R(t,\tau)$ and
$\overline{R}(t,\tau)$ should asymptote to it in the $t\to\infty$ and
$(\tau-t)\to \infty$ limits. We find that $R(t,\tau)$ and
$\overline{R}(t,\tau)$ overlap with the grey band, however the spread due
to remaining ESC is larger in $R(t,\tau)$ than in
$\overline{R}(t,\tau)$.  In fact, on the finer $a09m310$ ensemble, we
do not observe a spread in $\overline{R}(t,\tau)$ versus $\tau$, however,
the comparison with the grey band suggests that quoting the average of
the $\overline{R}(t,\tau)$ data as the final result could underestimate the error.

\begin{figure}[h]
  \center
  \includegraphics[width=0.49\textwidth]{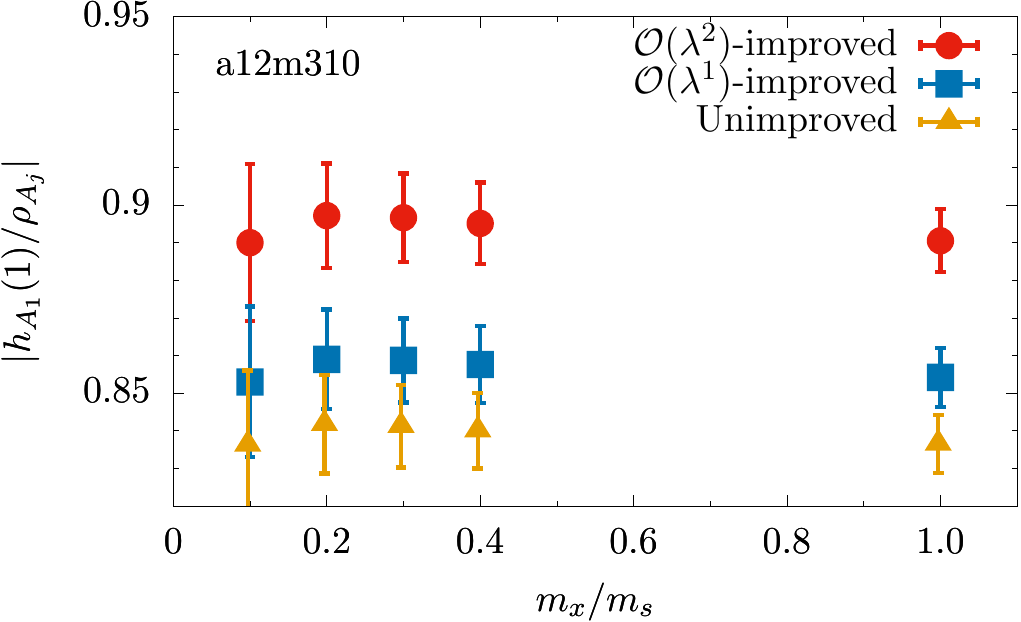}
  \includegraphics[width=0.49\textwidth]{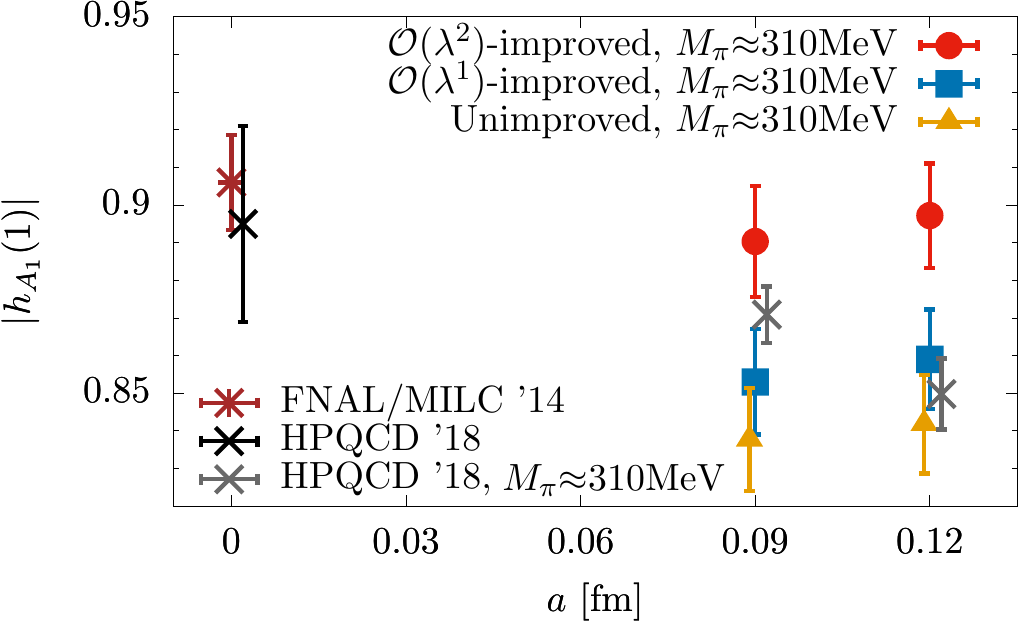}
  \caption{(Left) Data for $|h_{A_1}(1)/\rho_{A_1}|$ for three levels
    of current improvement plotted versus the spectator quark mass
    $m_x$ for the $a12m310$ ensemble. (Right) Comparison of data on
    two different ensembles, $a12m310$ and $a09m310$ at the unitary point $m_x=0.2m_s$, and with
    $\rho_{A_1}=1$ to investigate $a$ dependence, and to compare with
    FNAL/MILC and HPQCD collaboration results~\cite{Bailey:2014tva,
      Harrison:2017fmw} obtained in the continuum limit. }
  \label{fig:hA1}
\end{figure}

The data for $|h_{A_1}(1)/\rho_{A_j}|$ in Fig.~\ref{fig:hA1} (left)
show no significant dependence on the spectator quark mass $m_x$.  The
observed dependence on the order of improvement in the current is
unexpected from naive HQET power counting and could an artifact of
setting $\rho_{A_j}=1$, which also depends on the order of improvement. In
Fig.~\ref{fig:hA1} (right), we compare our results with those from the
FNAL/MILC and HPQCD collaborations~\cite{Bailey:2014tva,
  Harrison:2017fmw} obtained in the continuum limit. The
$O(\lambda^2)$ data are consistent with the FNAL/MILC and HPQCD
results and show no significant lattice spacing dependence. This rough
agreement provides a good and encouraging check of our calculations
that are being done with a much more complicated heavy quark action
and current.

A brief summary of the work under progress is as follows.  (i)
Analysis with the $\CO(\lambda^3)$-improvement terms in the current,
(ii) analysis of the data for the nonzero recoil form factors in $B\to
D^{(\ast)} \ell \nu$ decays, and (iii) the analysis for the decay
constants $f_D$, $f_{D_s}$, $f_B$, $f_{B_s}$ and $f_{B_c}$.

\acknowledgments
We thank the MILC collaboration for sharing the $2+1+1$-flavor HISQ
ensembles generated by them.  Computations for this work were carried
out in part on (i) facilities of the USQCD collaboration, which are funded
by the Office of Science of the U.S. Department of
Energy, and (ii) the DAVID GPU clusters at Seoul National University.
The research of W.~Lee is supported by the Creative Research
Initiatives Program (No.~2017013332) of the NRF grant funded by the
Korean government (MEST).
W.~Lee acknowledges support from the KISTI
supercomputing center through the strategic support program for the
supercomputing application research (No.~KSC-2016-C3-0072).

%

\bibliographystyle{JHEP}
\bibliography{ref} 

\end{document}